\documentclass[twocolumn]{pasj02}

\Received{}
\Accepted{}
 
\usepackage{xcolor}
\usepackage{natbib}

\usepackage[switch,mathlines]{lineno}

\usepackage{ulem}
\DeclareRobustCommand{\erase}{\bgroup\markoverwith{\textcolor{red}{\rule[.5ex]{2pt}{1.4pt}}}\ULon}
\DeclareRobustCommand{\eraseb}{\bgroup\markoverwith{\textcolor{blue}{\rule[.5ex]{2pt}{1.4pt}}}\ULon}

\begin{document} 

\title{ 
Detailed Chemical Abundance Analysis of Metal-Poor Turn-Off Stars: One with [Fe/H] $< -4$ and Three with [Fe/H] $< -3$.
}

\author{Takuma \textsc{Suda}\altaffilmark{1,2}%
\thanks{Example: Present Address is xxxxxxxxxx}}
\altaffiltext{1}{Department of Liberal Arts, Tokyo
University of Technology, 5-23-22 Kamata, Ota-ku, Tokyo 144-8535, Japan}
\altaffiltext{2}{Research Center for the Early Universe,
The University of Tokyo, 7-3-1 Hongo, Bunkyo-ku, Tokyo 113-0033,
Japan}
\email{sudatkm@stf.teu.ac.jp}

\author{Patrick \textsc{Fran\c cois}\altaffilmark{3, 4}}
\altaffiltext{3}{ LIRA, Observatoire de Paris, Universit{\'e} PSL, Sorbonne Universit{\'e}, Universit{\'e} Paris CitUniversit{\'e}, CY Cergy Paris Universit{\'e}, CNRS,75014 Paris, France
}
\altaffiltext{4}{UPJV, Universit\'e de Picardie Jules Verne, 33 rue St Leu, 80080 Amiens, France }

\author{Shinya \textsc{Wanajo}\altaffilmark{5}}
\altaffiltext{5}{Department of Astronomy, Faculty of Science, Tohoku University, Sendai, Miyagi 980-8578, Japan}

\author{Elisabetta \textsc{ Caffau}\altaffilmark{6}}
\altaffiltext{6}{ LIRA, Observatoire de Paris, Universit{\'e} PSL, Sorbonne Universit{\'e}, Universit{\'e} Paris CitUniversit{\'e}, CY Cergy Paris Universit{\'e}, CNRS,92195 Meudon, France}

\author{Wako \textsc{Aoki}\altaffilmark{7}}
\altaffiltext{7}{National Observatory of Japan, Mitaka, Tokyo, Japan}

\author{Piercarlo \textsc{Bonifacio}\altaffilmark{6}}



\maketitle

\begin{abstract}
We present the analysis of four new extremely metal poor turn-off stars thanks to high-resolution spectra obtained with the 
Subaru/HDS spectrograph. 
We determined the abundances and upper limits of Li, C, Mg, Ca, Sr, and Ba. 
Metallicities range from [Fe/H] $= -3.3$ to $-4.4$ dex.  For one of the stars, we measure the lithium abundance Log(Li/H) =2.1 $\pm$ 0.2. Two stars of the sample have low [$\alpha$/Fe] abundances.  
The most metal-poor star of the sample with [Fe/H] = -4.42 dex has a high [Sr/Fe] abundance ratio [Sr/Fe] = 0.9 dex, a high value also found in HE 1327-2326.
This star is the second-most iron-poor star observed with Subaru telescope, suggesting that more ultra metal-poor stars could be discovered using high-resolution spectrographs in the Northern Hemisphere.
\end{abstract}


\section{Introduction}

Extremely metal-poor (EMP)  stars are the witnesses of the early galactic chemical evolution.
Their chemical composition and their kinematical properties can be used to constrain the models of formation of the galaxies.
Numerical simulations suggest that the distribution of possible masses of the first stars may be much broader than previously believed, and may even extend down to a solar mass or below \citep{wollenberg_formation_2020,susa_merge_2019} leading to stars which are still alive and observable today.
The observed EMP stars are probably first or second generation stars.
In the former case their atmospheres must have been polluted \citep{yoshii1981,caffau2024}.
Their chemical composition provides a strong constraint on nucleosynthesis in the early Universe because EMP stars are the relics of the early chemical evolution.
The detailed chemical composition can be used to constrain the mass range and the type of supernova that enriched the matter of the star we observe now.

One of the main difficulties is to find these rare EMP among the other stars.
The traditional approach to identifying EMP stars involves examining a large number of stellar spectra, focusing on those that exhibit weak calcium doublet (H \& K) lines in the strong UV region.
A non exhaustive list includes the Hamburg ESO Survey (HES) \citep{christlieb_stellar_2008},  the use of Sloan Digital Sky Survey 
(hereafter SDSS, \citealt{SDSS12} )
spectra \citep{ludwig_extremely_2008}, and, more recently, the Pristine survey \citep{starkenburg_pristine_2017}.
In this article, we present the analysis of four stars that were identified in the SDSS DR16 using the calibration developed by \citet{ludwig_extremely_2008} . 
This article is a follow-up of  the Turn-Off Primordial Stars survey (TOPoS) that has been conducted as an ESO Large Programme at the VLT \citep{caffau_topos_2013}. One of the peculiarity  of this survey and one of its main advantages  was the ability to select metal-poor turn-off stars, giving the possibility to measure not only  the lithium abundance, but also the chemical composition of the stars unaffected by mixing.

In this study, we use two classification, EMP and UMP (ultra metal-poor), to categorize iron-poor stars.
These definitions are not rigid and are based on the origin of carbon-enhanced metal-poor stars (CEMP), which are characterised by [C/Fe] $\gtrsim 1$.
The typical range of iron-abundances for EMP stars is $-4 \lesssim$ [Fe/H] $\lesssim -3$, while for UMP stars, it is [Fe/H] $\lesssim -4$.
These criteria differ from those used in other studies, such as \citet{Beers2005}, both in terms of metallicity range and the conceptual framework for classification.
The classification of metal-poor stars are discussed below.


\section{Observations}

\begin{table*}
 \caption{Observation log } 
\label{tab:obslog}
\centering
\begin{tabular}{l c c c  }
\hline\hline
  Object          &  g magnitude  &  Observation date &  Exp. time  [s]    \\      
\hline

SDSS~J095932.52$+$265358.6  & 17.1 &	2023-04-10T07:35     & 3600.0  	\\  
SDSS~J095932.52$+$265358.6  & 17.1 &	2023-04-10T08:42	 & 3600.0 \\ 
SDSS~J095932.52$+$265358.6  & 17.1 &	2023-04-10T09:43	 & 3600.0 \\ 
SDSS~J161956.33$+$170539.9  & 17.8 &	2023-04-10T10:59	 & 3600.0  \\  
SDSS~J161956.33$+$170539.9  & 17.8 & 	2023-04-10T12:00	 & 3600.0  \\ 
SDSS~J161956.33$+$170539.9  & 17.8 & 	2023-04-10T13:01	 & 3282.0  \\ 
SDSS~J161956.33$+$170539.9  & 17.8 &	2023-04-10T13:59	 & 3600.0  \\ 
SDSS~J161956.33$+$170539.9  & 17.8 &	2023-04-10T14:53	 & 1803.0  \\ 
SDSS~J125601.88$+$460836.8  & 17.4 &	2023-06-10T06:29     & 3600.0  \\   
SDSS~J125601.88$+$460836.8  & 17.4 &	2023-06-10T08:14     & 3600.0  \\ 
SDSS~J125601.88$+$460836.8  & 17.4 &	2023-06-10T09:16     & 3600.0  \\
SDSS~J125601.88$+$460836.8  & 17.4 &	2023-06-10T10:17     & 3600.0 \\
SDSS~J165317.41$+$253728.4 & 17.38 &	2023-06-10T12:01     & 3600.0  \\
SDSS~J165317.41$+$253728.4 & 17.38 &	2023-06-10T13:06     & 3600.0  \\
SDSS~J165317.41$+$253728.4 & 17.38 &	2023-06-10T14:07     & 3600.0  \\

  \hline
  \end{tabular}
   \end{table*}

The observations were carried out with the High Dispersion Spectrograph (HDS) installed on the Subaru telescope \citep{noguchi_high_2002}.
The wavelength coverage goes from 4084 \AA ~to  6892 \AA. 
A binning $2 \times 2$ was adopted, leading to a resolving power of about 40\,000.
The logbook of the observations is given in table \ref{tab:obslog}.
Standard data reduction procedures were carried out with the IRAF Echelle package \footnote{ IRAF is distributed by the National Optical Astronomy Observatories, which is operated by the Association of Universities for Research in Astronomy, Inc. under cooperative agreement with the National Science Foundation}. Care was taken to remove the sky background as most of the exposures were affected by the moon illumination. 
\

\section{Stellar parameters}

 \begin{table*}
 \caption{Adopted stellar parameters for the list of targets. The second-to-last column lists the measured radial velocity of the stars after correction for the barycentric velocity.}
\label{tab:stellar_parameters}
\centering
\begin{tabular}{l lc c c c c c}
\hline\hline
  Object          & Gaia DR3 ID &  ${\rm T}_{eff}$ & log~g  & [Fe/H]   & $\xi$ (km/s)  &  $v_{r}$ (km/s) & S/N @ 480 nm \\      
\hline

SDSS~J095932.52$+$265358.6  & 740011079710836352  & 5491   &  4.68    &   -2.7    & 1.5    &    151 & 23\\   
SDSS~J125601.88$+$460836.8  & 1530789020754458624 & 5830   &  4.54    &   -2.98   & 1.5    &   23 & 19 \\      
SDSS~J161956.33$+$170539.9  & 4466828183263855872 & 6195   &  4.68     &   -4.00   & 1.5    &     -307 & 23  \\  
SDSS~J165317.41$+$253728.4  & 1306161917639679744 & 6210   &  3.97    &   -2.8    & 1.5    &      -193  & 20\\

  \hline
  \end{tabular}
   \end{table*}

\begin{figure}
\includegraphics[width=8cm]{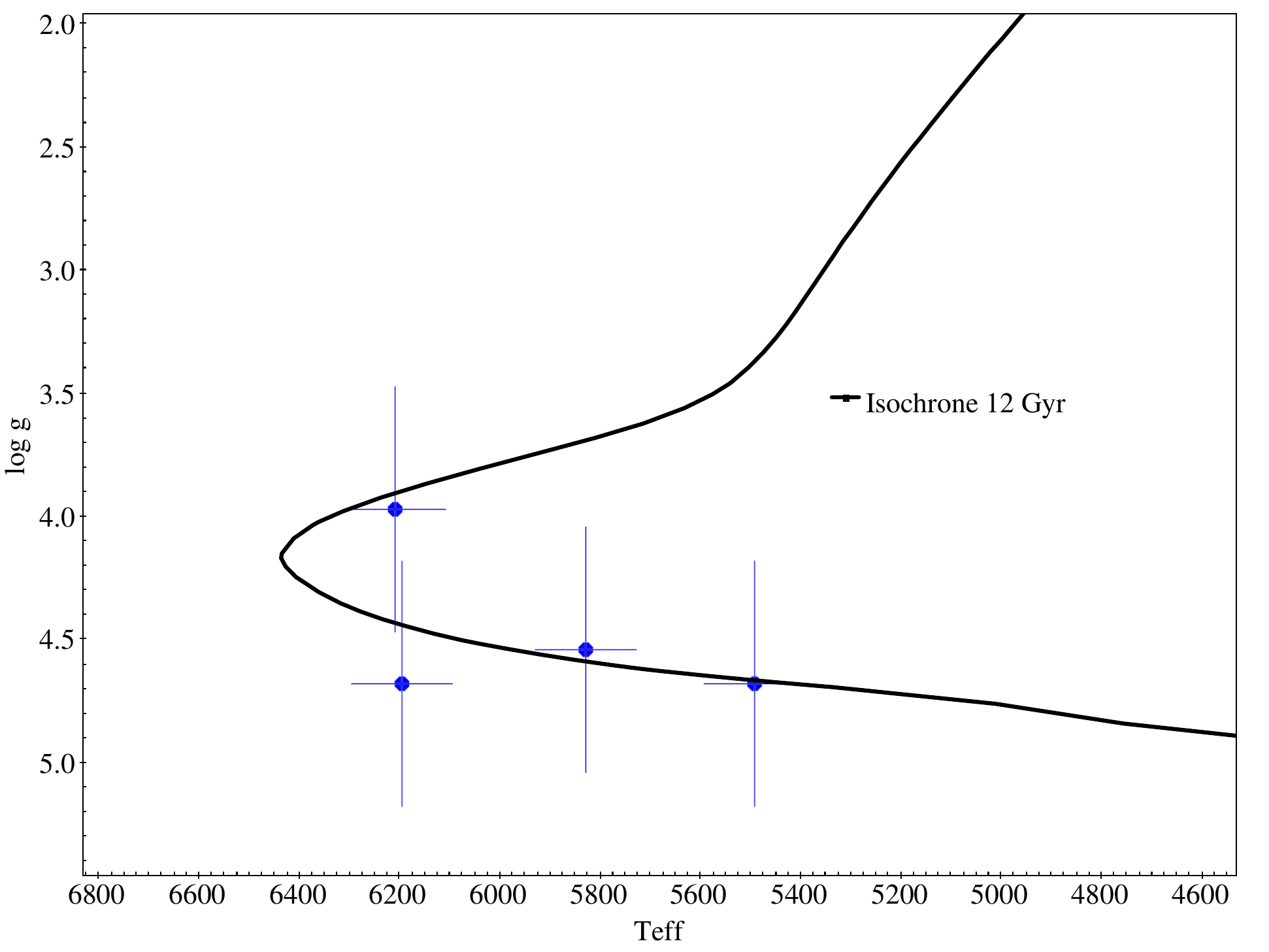} 
\caption{Comparison of the positions of the program stars and a PARSEC \citep{bressan_span_2012} isochrone with metallicity 
[Fe/H] $= -2.2$\ and age 12 Gyr on a Kiel diagram (log~g vs. $T_{eff}$).}
\label{fig:Kiel}
\end{figure}

All four stars are contained in the spectroscopic sample of \citet{bonifacio_topos_2021}\footnote{ http://vizier.cds.unistra.fr/viz-bin/VizieR?-source=J/A+A/651/A79} and the atmospheric parameters are provided in Table \ref{tab:stellar_parameters}.
We briefly recall how those parameters were derived.
The $g-z$ colour was corrected for reddening using the \citet{schlegel_maps_1998} extinction maps and the $(g-z)_{0}$ -- effective temperature calibration provided in \citet{bonifacio_topos_2021}.
The metallicities were derived using the code MyGisFoS \citep{sbordone_mygisfos_2014} to analyse the SDSS spectra of the stars SDSS~J095932.52$+$265358.6, SDSS~J125601.88$+$460836.8 and SDSS~J161956.33$+$170539.9 and the BOSS spectrum for the star SDSS~J165317.41$+$253728.4, assuming log g = 4.0.
The surface gravities in Table \ref{tab:stellar_parameters} were derived from the Stefan Boltzman equation, assuming the above derived effective temperatures, the Gaia $G$ magnitude, the distance estimates derived by \citep{2018AJ....156...58B} from the Gaia parallaxes and a mass of $0.8 M_\odot$. 
We have also plotted in our stars in a Kiel diagramme (Fig.~\ref{fig:Kiel}) together with a PARSEC (PAdova and TRieste Stellar Evolution Code) \citep{bressan_span_2012} isochrone corresponding to the metallicity of [Fe/H] $= -2.2$ and an age of 12 Gyr. The location of the stars clearly reveals that the stars are dwarfs near the turn-off, with one star exhibiting mild post turn-off evolution.

\begin{figure}
\includegraphics[width=8cm]{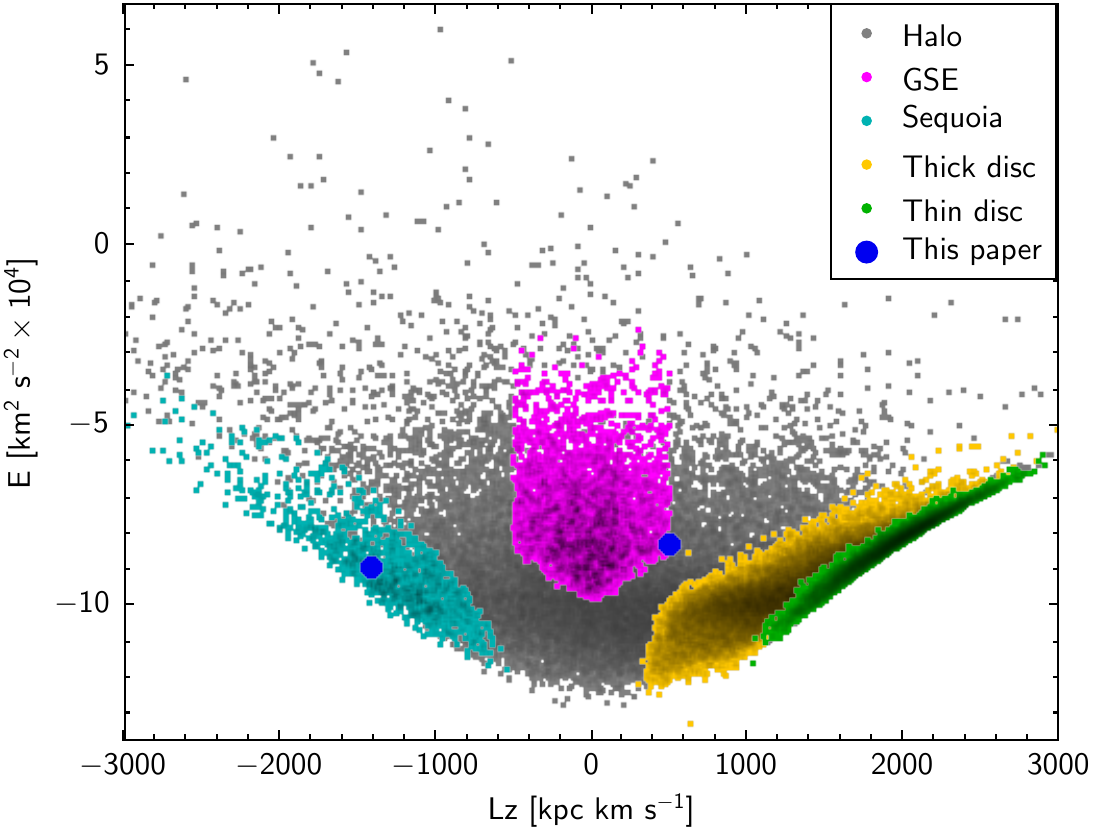} 
\caption{The specific energy versus specific  angular plot for the ``good parallax''
sample of \citep{bonifacio_topos_2021}.
The stars SDSS\,J095932.52+265358.6 and SDSS\,J125601.88+460836.8
are shown as a blue $X$ symbol. The GSE and Sequoia events selected
according to the criteria of  \citep{feuillet2021}.
The thin and thick discs are selected according to the criteria of \citet{bensby14}.
{Alt text: Dot graph showing the kinematic properties of the Galactic stars.}
}
\label{fig:elz}
\end{figure}

One more remark of our four targets only two have Galactic kinematics computed by \citet{bonifacio_topos_2021}: SDSS~J095932.52$+$265358.6 and SDSS~J125601.88$+$460836.8 (see Figure~\ref{fig:elz}).  
The former has a kinematics marginally consistent with the accretion event GSE \citep{2018MNRAS.478..611B,2018ApJ...863..113H,2018Natur.563...85H}, the latter is on a retrograde orbit and is consistent with belonging to the Sequoia/Thamnos accretion event  \citep{barba19,myeong19,villanova19,koppelman19}.

\section{Analysis}

We carried out a classical 1D LTE analysis using OSMARCS model atmospheres  \citep{gustafsson_grid_1975,   gustafsson_grid_2003,   gustafsson_grid_2008,  plez_spherical_1992, edvardsson_chemical_1993}.
The abundances used in the model atmospheres were solar-scaled with respect to the \citet{Grevesse2000} solar abundances, except for the $\alpha$ elements that are enhanced by 0.4 dex.
Corrections on the resulting abundances are considered to take into account the difference  between  \citet{Grevesse2000} and  \citet{caffau_solar_2010}, \citet{lodders_abundances_2009} solar abundances.
Finally, the solar abundances adopted for this work are log(C/H)$_{\odot}$=8.50,     log(Mg/H)$_{\odot}$=7.54,   log(Ca/H)$_{\odot}$=6.33, log(Fe/H)$_{\odot}$=7.52, log(Sr/H)$_{\odot}$=2.92 and log(Ba/H)$_{\odot}$=2.17. 

The abundance analysis was performed using the code turbospectrum, a LTE spectral line analysis code developped by \citet{alvarez_near-infrared_1998} and \citet{plez_turbospectrum_2012}, which treats scattering in detail.
The carbon abundance was determined by fitting the CH band near to 430 nm (G-band).
The molecular data that correspond to the CH band are described in \citet{hill_extreme_2002}.
The abundances are determined by matching a synthetic spectrum centred on each line of interest to the observed spectrum.
Table \ref{tab:linelist} presents the list of spectral lines used to measure abundances or determine upper limits in our sample of stars.

\begin{table*}
 \caption{List of  absorption lines used to determine the abundances.} 
\label{tab:linelist}
\centering
\begin{tabular}{l r r r }
\hline\hline
 Element &  Wavelength ({\rm$\AA$})   & \rm{$\chi_{esc}$} & log~gf \\
 
 \hline
 LiI         &   6707.761   & 0.00   &  -0.009  \\
 LiI         &   6707.912   & 0.00   &  -0.309  \\
 CH band     &   4315       &        &          \\
 CH band     &   4324       &        &          \\
 MgI         &   5172.698   & 2.71   &  -0.38   \\
 MgI         &   5183.619   & 2.72   &  -0.16   \\
CaI          &   4226.740   & 0.00   &  +0.24   \\   
FeI          &   4202.040   & 1.48   &  -0.70   \\
FeI          &   4260.486   & 2.40   &  -0.02   \\
FeI          &   4271.164   & 2.45   &  -0.35   \\
FeI          &   4325.775   & 1.61   &  -0.01   \\
FeI          &   4383.557   & 1.48   &   0.20   \\
FeI          &   4404.761   & 1.56   &  -0.14   \\
FeI          &   4415.135   & 1.61   &  -0.61   \\
FeI          &   5269.550   & 0.86   &  -1.32   \\
SrII         &   4215.520   & 0.00   &  -0.17   \\  
BaII         &   4554.036   & 0.00   &  +0.16   \\
   \hline
  \end{tabular}
   \end{table*}

\section{Errors}\label{sec:error}

\begin{table*}
 \caption{ Estimated errors in the element abundance ratios [Fe/H] and  [X/Fe] for the star  SDSS~J095932.52$+$265358.6. The other stars give similar results. } 
\label{tab:errors}
\centering
\begin{tabular}{l c c c }
\hline\hline
  Element   &         $\Delta T_{eff}$ =  100~K  & $\Delta$ log~g = 0.5~dex   &$\Delta$  $ v_{t}$ =   0.5~km/s \\     
\hline
$[$C/Fe$]$    &   0.2  &   0.2  &     0.1   \\
$[$Mg/Fe$]$ &   0.1  &  0.15 &  0.15  \\
$[$Ca/Fe$]$ &   0.1  &  0.1  &  0.15 \\  
$[$Fe/H$]$   &   0.1 &   0.04  &     0.1    \\
$[$Sr/Fe$]$ &   0.1  &  0.25  &  0.25  \\
$[$Ba/Fe$]$ &   0.1  &  0.25   &  0.3  \\
  \hline
  \end{tabular}
   \end{table*}

Table \ref{tab:errors} lists the computed errors in the elemental abundance ratios due to typical uncertainties in the stellar parameters.
The errors were estimated by varying  $T_{eff}$ by $\pm$ 100~K, log~g  by $\pm$  0.5~dex, and $ v_{t}$  by $\pm$ 0.5 dex in the model atmosphere of SDSS~J095932.52$+$265358.6.
  These are the typical uncertainties used to estimate the sensitivity of each parameter on the abundance determination. These adopted values for   $T_{eff}$ and $ log g $ are of the order of the standard deviation  that can be found  between the stellar parameters derived using MyGisFos  \citep{sbordone_mygisfos_2014} and the stellar parameters obtained by Starhorse \citep{anders_photo-astrometric_2019}.


In this star, we could measure the Mg, Ca, Fe  and set limits for Li, Sr, and Ba  abundances.
The main uncertainty comes from the error in the placement of the continuum when the synthetic line profiles are matched to the observed spectra.
In particular, residuals from the sky subtraction may lead to a decrease of the S/N ratio.
Since the final spectra are built by combining multiple exposures taken at different epochs, each with different barycentric velocities, residual sky features are smoothed, which may degrade the S/N of the spectra.

All the abundances have  been computed using spectrum synthesis. The error associated to the fit is a combination of the quality of the  fit and the error on the continuum placement. The estimate of these errors is evaluated by executing multiple line synthesis fitting with different assumption for  the continuum position. 
The associated error ranges from approximately 0.1 to 0.2, depending on the S/N ratio of the spectrum and the species under consideration, with the largest uncertainty observed for neutron capture elements.
Finally,  uncertainties on the stellar parameters and the errors on the line fitting can be summed quadratically.


We estimated the NLTE corrections using the database provided by \citet{mashonkina_1d_2023}\footnote{ https://spectrum.inasan.ru/nLTE2/}, which offers a web-based interface for interpolating 1D NLTE corrections from the literature. Although the correction for a given element can reach up to $\simeq$ 0.2 dex, the corrections for abundance ratios [X/Fe] are generally below 0.15 dex, and in most cases, below 0.05 dex. 

The use of resonance lines in our abundance analysis is justified, as discussed for the calcium line in the Appendix. Owing to the limited availability of absorption lines in EMP dwarfs, our analysis relies on a different set of lines than those used in studies of brighter or more metal-rich stars. The difference between abundances derived from the Ca~I resonance line and those from the Ca~II HK lines or the Ca~II triplet can reach up to $\sim$ 0.5 dex at [Fe/H] $\lesssim -5$, but is typically less than 0.1 to 0.2 dex at [Fe/H] $\sim -3$.

\section{Abundance results and discussion}

The abundances and upper limits for the sample of stars in this programme are compiled in Table \ref{tab:ratios}. For lithium and carbon, the log(X/H) is given  whereas [X/Fe] results are presented for magnesium, calcium, strontium and barium.
For the star SDSS~J161956.33$+$170539.9, we found a metallicity of [Fe/H] = -4.42 dex making it a new ultra-metal-poor (UMP) star. 
This star is the second most iron-poor star observed with Subaru following HE 1327-2326, which has [Fe/H] = -5.6 \citep{Aoki2006}.

\begin{figure}[h!]
 \begin{center}
\includegraphics[width=8cm]{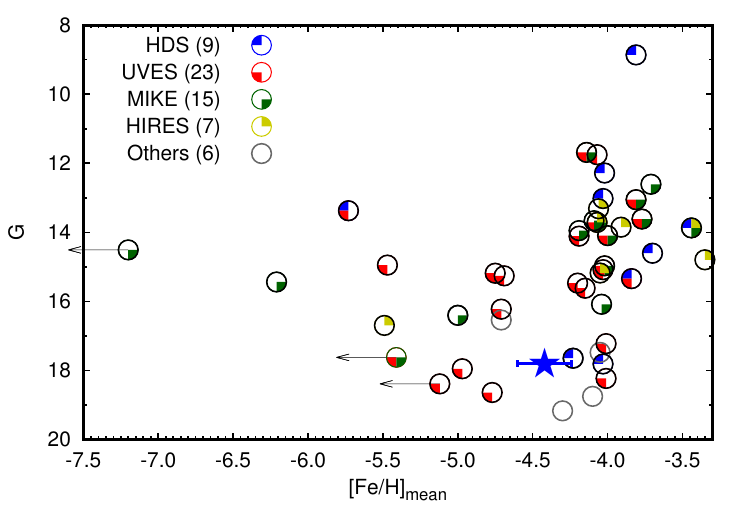} 
 \end{center}
\caption{
G-band magnitude as a function of mean metallicity for stars with [Fe/H] $\lesssim -4$. The data are taken from the SAGA database, where stars are selected based on at least one paper reporting [Fe/H] $\leq -4$. G-band photometry is obtained from Gaia DR3 via the SIMBAD database. Arrows for metallicity indicate upper limits, where the average is calculated for data with only upper limits available. For stars with measured [Fe/H], data with upper limits are excluded when computing the mean [Fe/H]. The star symbol represents our data for SDSS~J161956.33$+$170539.9. Other symbols, shown as partially filled circles, indicate the facilities used to obtain spectra for metallicity determination. The upper filled circles represent telescopes in the Northern Hemisphere (i.e., Subaru and Keck), while the lower filled circles correspond to those in the Southern Hemisphere (i.e., VLT and Magellan). Black circles denote stars observed with other telescopes. The names of the corresponding spectrographs (Ultraviolet and Visual Echelle Spectrograph for UVES, the Magellan Inamori Kyocera Echelle for MIKE, and High Resolution Echelle Spectrometer for HIRES) and the number of stars are provided in parentheses in the top-left corner of the panel.
{Alt text: Dot graph showing the relationship between brightness and metallicity of the most iron-poor stars in the Milky Way halo.}
}\label{fig:ump}
\end{figure}

Figure~\ref{fig:ump} presents the known UMP stars with [Fe/H] $\lesssim -4$, taken from the SAGA database \citep{Suda2008,Suda2011,Yamada2013,Suda2017}. The [Fe/H] values represent the average of all reported metallicities, excluding upper limits. When only upper limits are available, we calculate the average of these values, which are indicated by arrows. It is to be noted that there may be a bias against reserch groups to calculate the mean metallicity because the same research group sometimes share the previous results, which are compiled as separate data in the database. Some stars with [Fe/H] $> -4$ are plotted because the data are included if at least one paper reports [Fe/H] $\leq -4$. Most of these stars are likely EMP stars rather than UMP stars, as only \citet{roederer_search_2014} reports [Fe/H] $\leq -4$, while other papers do not.
Error bars are omitted in Fig. 2 and subsequent figures to avoid confusion arising from different treatments of uncertainties and to maintain the clarity of the plots.

The figure also depicts the facilities used to measure the iron abundances of UMP stars, represented by partially filled circles. Circles filled in the upper half indicate observations made with the Subaru telescope or the Keck I telescope, which are located in the Northern Hemisphere, while circles filled in the lower half represent observations conducted with the Very Large Telescope (VLT) or the Magellan telescopes in the Southern Hemisphere.
Some of the stars are observed with multiple facilities including those not listed in the figure. These facilities include X-Shooter on VLT, the Intermediate dispersion Spectrograph and Imaging System on the 4.2 m William Herschel Telescope, and the Optical System for Imaging and low-intermediate-Resolution Integrated Spectroscopy instrument on the 10.4 m Gran Telescopio Canarias telescope. These instruments have also contributed to the discovery of UMP stars.

From the figure, we can expect more UMP stars to be visible in the Northern Hemisphere. Currently, only two out of 14 stars with [Fe/H] $< -4.5$ are discovered using large ground-based telescopes in the Northern Hemisphere. It is also likely that stars with [Fe/H] $< -5$ could be found among relatively bright targets (G $\sim$ 15 mag). Future facilities, such as $\sim$30-meter-class telescopes, will greatly enhance the search for stars with [Fe/H] $< -5$ and G $\gtrsim 18$.

 \begin{table*}[h!]
 \caption{Lithium and  carbon abundances. $\alpha$ and  neutron-capture element  abundance ratios. Therefore, the abundances are given  as A(X) for Li and C  and  [X/H] for Fe
 and [X/Fe] for Mg, Ca, Sr and Ba.}   
\label{tab:ratios}
\centering
\begin{tabular}{l r r r r r r r  }
\hline\hline
            Object           &  [Fe/H]   &      A(Li)    &         A(C)   & [Mg/Fe]   &        [Ca/Fe] &       [Sr/Fe]    & [Ba/Fe]        \\    
 \hline 
SDSS~J095932.52$+$265358.6   &  -3.52$\pm$0.18    & $\le$ 1.80    &  $\le$ 6.0     &   0.58$\pm$0.25    &         0.59$\pm$0.23   &         -0.10$\pm$0.42    & $\le$  0.35   \\    
SDSS~J125601.88$+$460836.8   &  -3.39$\pm$0.18    &       2.10$\pm$0.11    &  $\le$ 7.0     &   0.10$\pm$0.25    &        -0.14$\pm$0.23   &  $\le$  -0.23    & $\le$   0.22   \\         
SDSS~J161956.33$+$170539.9   &  -4.42$\pm$0.18    & $\le$ 2.10    &  $\le$ 7.6     &   0.45$\pm$0.25    &         0.59$\pm$0.23   &          0.90$\pm$0.42    & $\le$  1.05   \\     
SDSS~J165317.41$+$253728.4   &  -3.77$\pm$0.18    & $\le$ 2.10    &  $\le$ 7.45    &  -0.02$\pm$0.25    &         0.24$\pm$0.23   &  $\le$  -0.25    &  $\le$ 0.80   \\    
 \hline\hline
\end{tabular}
\end{table*}

\subsection{Lithium}

\begin{figure}[h!]
 \begin{center}
\includegraphics[width=8cm]{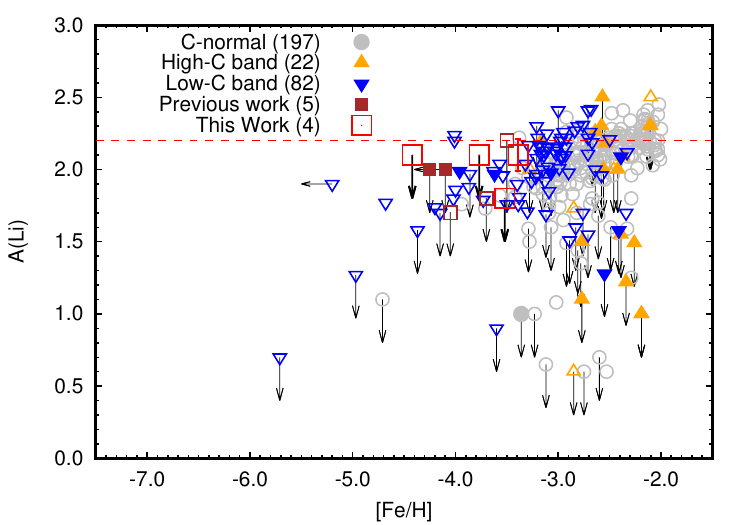} 
 \end{center}
\caption{Lithium abundance in unevolved EMP stars with [Fe/H] $< -2$. The red squares represent our results. The brown squares refer to our previous analysis \citep{francois_detailed_2020} of Subaru HDS spectra. The circles represent carbon-normal stars selected from the literature using the SAGA database. CEMP stars of the low- and high-carbon bands are shown by orange upper triangles and blue lower triangles, respectively. Upper limits are shown by arrows.
Filled symbols represent Ba-enhanced stars with [Ba/Fe] $> 1$.
Open symbols represent the stars with low abundances for neutron-capture elements or those abundances are not available for the literature data. The red dashed line represents the Spite plateau as determined by \citet{sbordone_metal-poor_2010}. The numbers in parentheses in the top-left corner indicate the number of stars in each group. Details of the symbols and the criteria for the CEMP sub-classes are described in the text.
{Alt text: Dot graph.}
}\label{fig:lithium}
\end{figure}

\begin{figure}[h!]
 \begin{center}
\includegraphics[width=8cm]{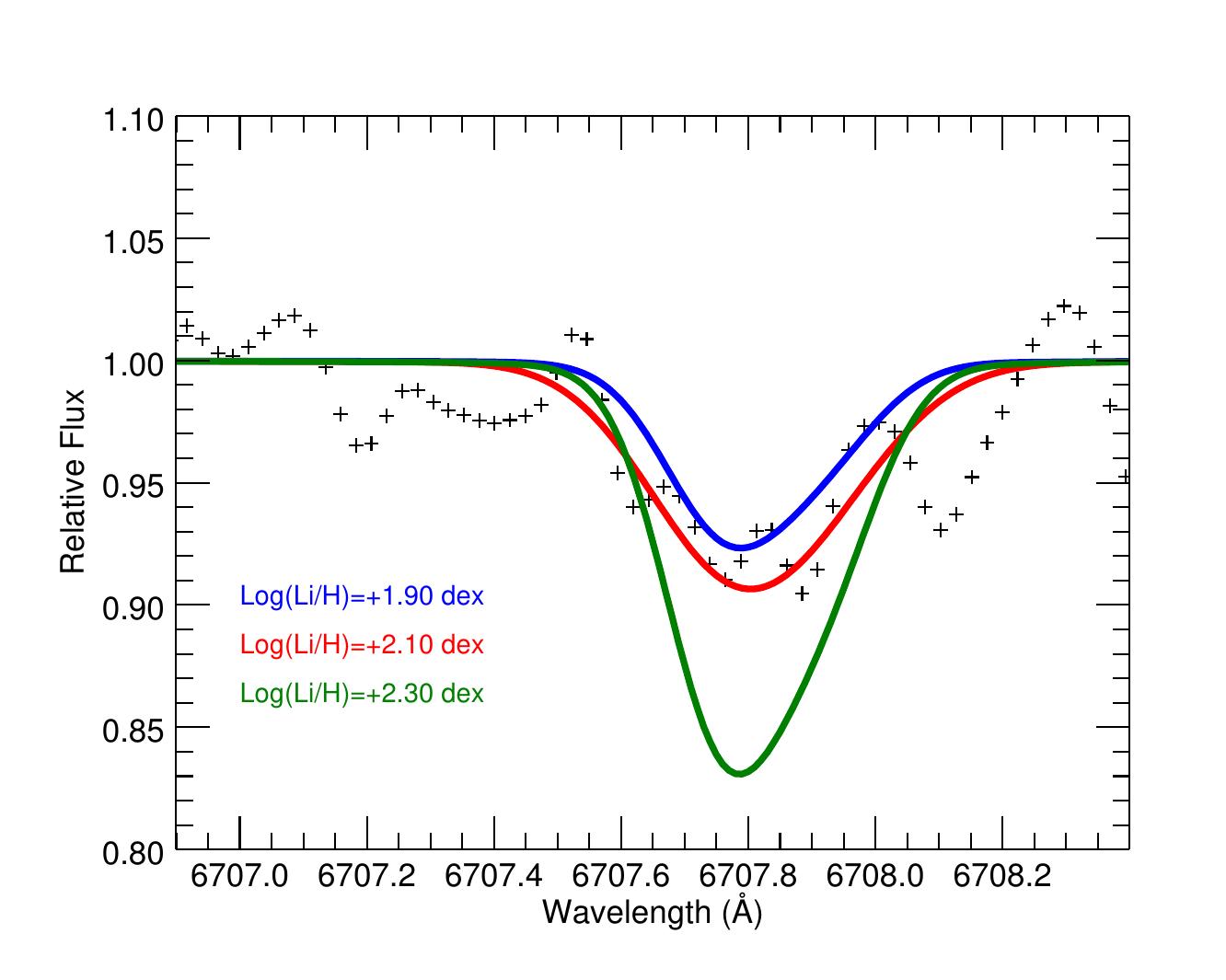} 
 \end{center}
\caption{ Comparison of the observed spectrum of  SDSS~J125601.88$+$460836.8  with the synthetic spectrum. The observed spectrum is shown as '+' symbols. The theoretical spectrum with Log(Li) = 2.1 dex is represented as a red line.
{Alt text: One line graph with dots showing the fit of a spectral synthesis model to the observed spectrum.}
}
\label{fig:lithium_fit}
\end{figure}

 Fig. \ref{fig:lithium} shows our lithium abundance results represented as red squares.
As lithium is diluted in the convective envelopes of red giants, we made the comparison with the lithium abundances determined in unevolved stars from the literature using the SAGA database.
The selection criteria for unevolved stars include [Fe/H] $< -2$ and $\log g \geq 3.5$, following the stellar evolution models of 0.8 $M_{\odot}$ at very low metallicity \citep{Suda2011}. If lithium abundance is reported in multiple papers, we have arbitrarily selected one data point, considering factors such as spectral resolution, abundance constraints, the number of measured elements, and other relevant criteria
\citep{frebel_he_2008,MatasPinto2021,plez_analysis_2005,Hansen2011,Lardo2021,SiqueiraMello2015,Lai2008,caffau_topos_2016,caffau_extremely_2011,Li2018,Li2022,thompson_cs_2008,roederer_search_2014,Jeong2023,Spite2015,Spite2022,bonifacio_topos_2015,bonifacio_topos_2018,lucatello_stellar_2003,hansen_elemental_2015,masseron_lithium_2012,Matsuno2017AJ,matsuno_lithium_2017,Sivarani2004,sivarani_first_2006,placco_g64-12_2016,aoki_carbon-enhanced_2008,Aoki2010}.
We obtained a detection of lithium in the star  SDSS~J125601.88$+$460836.8  with a value A(Li) = 2.1 dex. 
The comparison of the observed spectrum with the synthetic spectrum is shown in Fig. \ref{fig:lithium_fit}.

In Fig.~\ref{fig:lithium}, carbon-normal stars are represented by gray circles, while high- and low-carbon-band stars are shown as upper and lower triangles, respectively.

We followed the classification of CEMP stars proposed by \cite{bonifacio_topos_2018}:

 \begin{itemize}
 
 \item{} Carbon-normal stars: For [Fe/H] $\geq$ -4.0, [C/Fe] $<$ 1.0; for [Fe/H] $<$ -4.0 dex, A(C) $<$ 5.5.;
 
  \item{} Low-carbon band CEMP stars: Stars that do not meet the carbon-normal criterion and have A(C) $\leq$ 7.6.
  
   \item{} High-carbon band CEMP stars: Stars that do not meet the carbon-normal criterion and have A(C) $>$ 7.6.
   
 \end{itemize}

These boundaries are shown by the dashed lines in Figure~\ref{fig:carbon}.
Open symbols represent stars with low [n-capture/Fe], where [Sr/Fe] or [Ba/Fe] is likely below 1 dex based on upper limits.
Open squares are used for our data because the abundances of neutron-capture elements are not well constrained.
Ba-enhanced stars with [Ba/Fe] $> 1$, represented by filled symbols, constitute 20\% of CEMP stars in the figure. However, they are more frequently found in the high-carbon band, i.e., 7 \% among low-carbon band, and 77 \% among high-carbon band.

High-carbon band CEMP stars tend to be more lithium-depleted than low-carbon band stars and carbon-normal stars in the metallicity range of $-3 <$ [Fe/H] $< -2$.
 \citet{bonifacio_topos_2018} argued that the high-carbon-band CEMP stars are the result of mass transfer from an AGB companion.
This suggests that the surface composition of observed high-carbon-band stars has been altered by accreted matter, meaning that the envelopes of donor stars are largely depleted of lithium. Additionally, the thin surface convective zones of unevolved low-mass stars are either well-mixed or dominated by the accreted material.
This could have significant implications for the evolution of binary systems.

In our sample of four EMP stars, we evaluated the upper limit in three of them and could measure the lithium abundance in one of them. 
One of the star (SDSS~J095932.52$+$265358.6 ) has a lithium upper limit lower than the Spite plateau value (Spite \& Spite 1982; Bonifacio et al. 2007; Sbordone et al. 2010).
This star lies in  the region where the ``meltdown'' of the lithium plateau is present (Sbordone et al. 2010; Aoki et al. 2009; Bonifacio et al. 2007).
However, caution is required as this star lies on the lower main sequence, with $T_{\rm eff} = 5491$ K and $\log g = 4.68$.
Its mass is approximately $0.7 M_{\odot}$, based on stellar evolution models for [Fe/H] $= -3$ \citep{Suda2010}.
At 13 Gyr, the mass of the surface convective zone is estimated to be $\sim 0.01M_{\odot}$, which is more than ten times larger than that of $\sim 0.8M_{\odot}$ models, which lie on the turn-off point on the H-R diagram, at the same evolutionary stage.
This suggests that lithium depletion in this star occurs through a mechanism other than dilution, given the absence of carbon and {\it s}-process element enhancement.
For two of the remaining  stars, we found higher upper limits for the lithium abundance at the level of A(Li) $\simeq$ 2.1, which is not enough to confirm the ``meltdown''. This does not rule out the possibility that these two stars have a much lower lithium abundance. Finally, for the star SDSS~J125601.88$+$460836.8, we measured a lithium abundance of A(Li) =2.1 dex, which is slightly below the lithium plateau.

\subsection{Carbon}

\begin{figure}[h!]
 \begin{center}
\includegraphics[width=8.0cm]{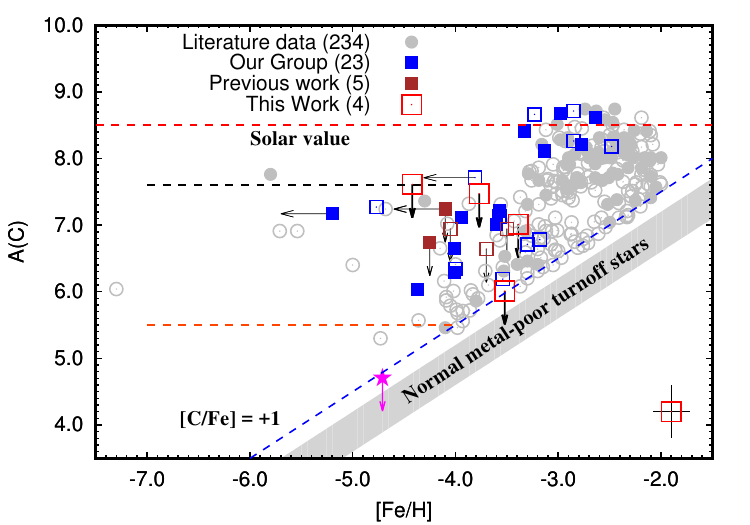} 
 \end{center}
\caption{Carbon abundances A(C) of CEMP stars as a function of [Fe/H], selected from the SAGA database with [Fe/H] $< -2$. Red squares represent our results. Arrows indicate the upper limits. Filled symbols represent Ba-enhanced stars with [Ba/Fe] $> 1$, while open symbols indicate Ba-normal stars or those without constraints on Ba abundances. Blue squares represent data published by our group, while data retrieved from the SAGA database are shown as gray circles.

The pink symbol represents SDSS~J102915$+$17292, the normal carbon ultra metal poor star discovered by \citet{caffau_extremely_2011}.
The black and orange dashed lines delimit the low-carbon band. Details can be found in \citet{bonifacio_topos_2018}.
The abundance uncertainties in this work are shown in the bottom right corner. See \S~\ref{sec:error} for details.
Uncertainties in the abundance measurements are indicated by an open square with error bars in the bottom right corner, and the same convention is used in the subsequent figures. For further details, see \S~\ref{sec:error}.
{Alt text: Dot graph showing carbon abundance as a function of metallicity.}
}
\label{fig:carbon}
\end{figure}

Figure \ref{fig:carbon} presents our results for carbon abundances, shown as red squares with arrows, as we obtained only upper limits for all target stars. 
In this plot, we have removed carbon-normal stars in Fig.~\ref{fig:lithium} and added CEMP giants from the literature: \citep{Arentsen2023,Frebel2007,Frebel2015,Goswami2006,Barbuy2005,Karinkuzhi2021,Karinkuzhi2017,Aguado2017,Aguado2018,Aguado2022,yong_most_2013,Yong2021,caffau_x-shooter_2013,Caffau2018,Depagne2002,Ito2013,Li2015ApJ,Li2015PASJ,Jacobson2015,Norris2007,cohen_normal_2013,Hollek2015,Shejeelammal2022,Rasmussen2020,Jonsell2006,Mashonkina2023,Gull2018,Gull2021,Mardini2019,Purandardas2019,Purandardas2021a,Purandardas2021b,Roriz2017,Francois2018,Goswami2021,Goswami2022,Keller2014,Schuler2007,hansen_elemental_2015,masseron_lithium_2012,Hansen2018,Placco2013,Placco2020,Aoki2002ApJ,Aoki2002aPASJ,Aoki2002bPASJ,Aoki2007,Aoki2018,Cui2013}.
Results from our previous analysis \citep{francois_detailed_2020} of Subaru HDS spectra are shown as brown squares, with open squares representing stars with low neutron-capture abundances. The pink star symbol represents SDSS~J102915$+$17292, the carbon-normal UMP star discovered by \citet{caffau_extremely_2011}.

Based on the upper limits of the carbon abundance in our four stars, we classify them as either moderately carbon-enhanced or carbon-normal. Since our measurements provide only upper limits, we cannot determine whether these stars are CEMP stars belonging to the low-carbon band or carbon-normal stars.

\subsection{Magnesium and calcium}

\begin{figure*}[h!]
 \begin{center}
\includegraphics[width=8cm]{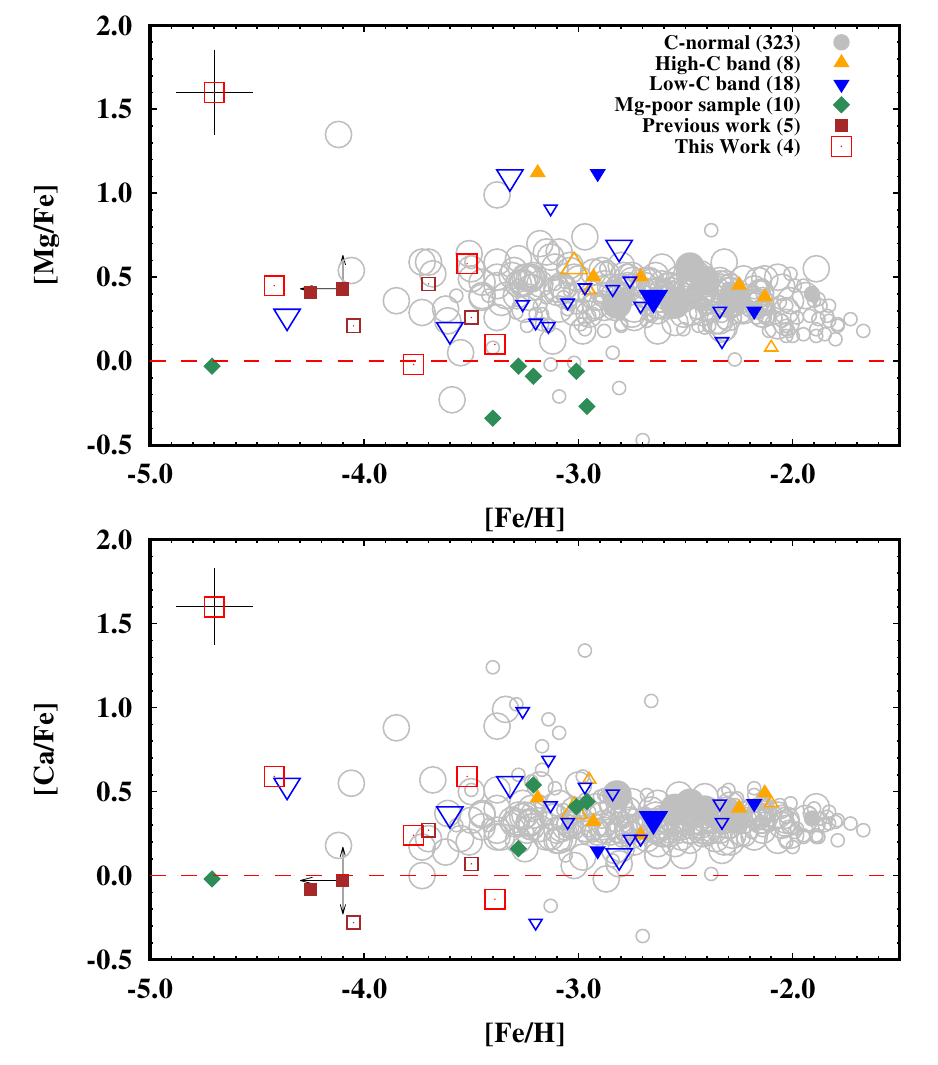} 
 \end{center}
\caption{[Mg/Fe] and [Ca/Fe] abundance ratios as a function of [Fe/H].
Red squares represent data from this paper. Grey circles indicate evolved and unevolved stars from \citet{Li2022}. Brown squares correspond to turn-off stars from \citet{francois_detailed_2020}. Green diamonds denote additional Mg-poor stars retrieved from the SAGA database.
Evolved stars are represented by large symbols, while unevolved stars are represented by small symbols.
Open and filled symbols have the same meanings as those in Fig.~\ref{fig:lithium}.
{Alt text: Two dot graphs showing [Mg/Fe] and [Ca/Fe] as a function of [Fe/H].}
}
\label{fig:alpha}
\end{figure*}

In Figure \ref{fig:alpha}, we present our results for Mg and Ca, the two $\alpha$-elements measurable in our spectra, as red squares. Literature data for dwarfs and giants are taken from the LAMOST-Subaru sample \citep{Li2022}, representing the combined data in Figs~\ref{fig:lithium} and \ref{fig:carbon}, which can be considered a homogeneous sample.

Among the four stars in our sample, two are $\alpha$-poor, with [Mg/Fe] ratios close to the solar value. Additionally, their [Ca/Fe] ratios are lower than the typical values found in most EMP stars. These results, based on high-resolution spectra, confirm the presence of $\alpha$-poor EMP stars, as previously suggested by \citet{bonifacio_topos_2018}.
The fraction of Mg-poor stars is approximately 1\% among metal-poor stars \citep{Li2022}. We also included known unevolved Mg-poor stars retrieved from the SAGA database \citep{caffau_extremely_2011,roederer_search_2014,hansen_elemental_2015,Francois2018,MatasPinto2021}, which make up around 1\% of the total sample.
Our results also suggest that the dispersion in the [$\alpha$/Fe] ratio increases as metallicity decreases, down to [Fe/H] $\sim -3.5$ for EMP stars. This trend is also observed in the sample of giant stars from \citet{Li2022}.

\begin{figure*}[h!]
    \begin{center}
  \includegraphics[width=8cm]{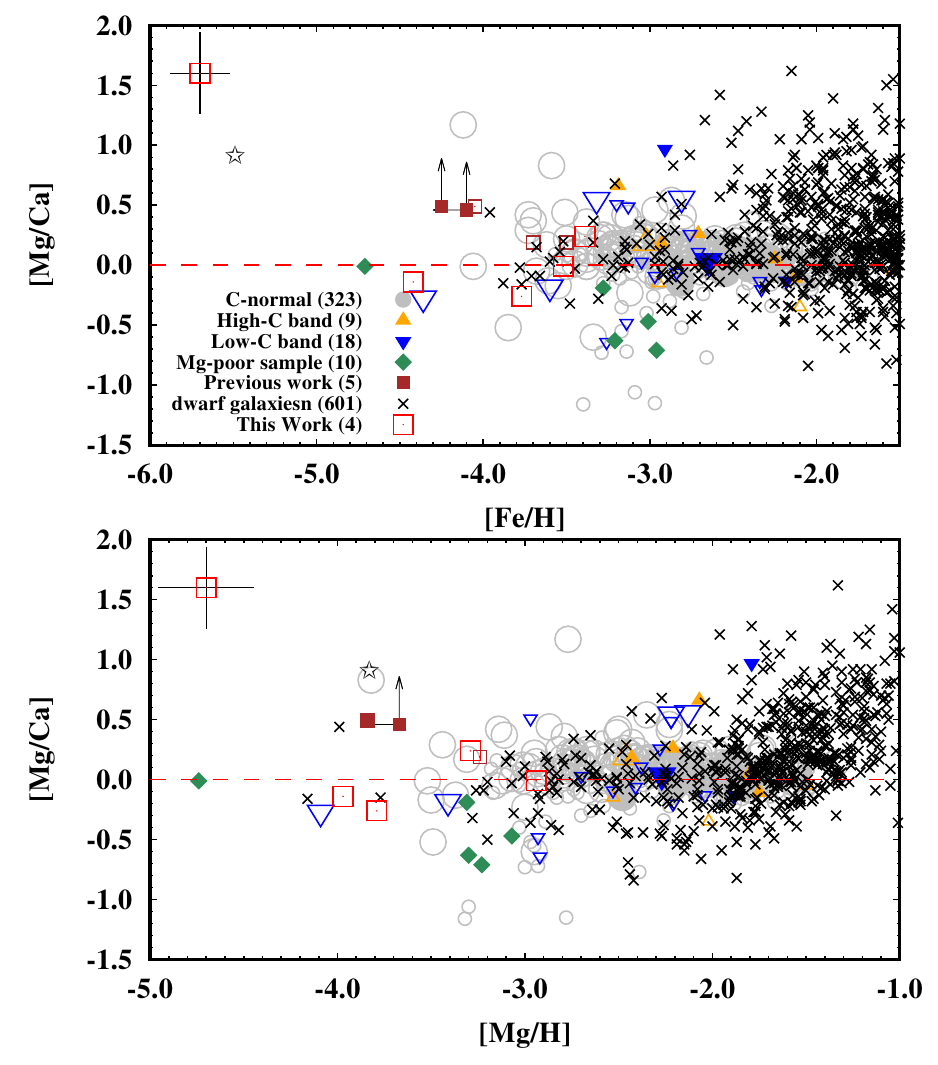}
   \end{center}
      \caption{[Mg/Ca] vs [Fe/H] and  [Mg/Ca] vs [Mg/H]. The meaning of the symbols is the same as in Fig.~\ref{fig:alpha}.
	  Crosses indicate stars in 26 dwarf galaxies in the local group, compiled from the SAGA database \citep{Suda2017}.
	  The star symbol marks SDSS~J081554.26$+$472947.5 \citep{gonzalez_hernandez_extreme_2020}, which shows an unusually high [Mg/Ca] at very low-metallicity.
               {Alt text: Two dot graphs showing [Mg/Ca] vs. [Fe/H] for the top panel and [Mg/Ca] vs. [Mg/H] for the bottom panel.}
              }
         \label{fig:MgCa}
   \end{figure*}

 Figure \ref{fig:MgCa} shows the [Mg/Ca] ratio as a function [Fe/H] (upper panel) and as a function of [Mg/H] (lower panel).
We added the literature data from \citet{Li2022} and Mg-poor stars from the SAGA database for comparison.
The [Mg/Ca] ratio found in our sample of stars ranges from $-0.2$ to $+0.2$ dex.
This ratio corresponds to  the abundance ratio found in the [Fe/H] range of $-2$ to $-3$ dex.  
At lower metallicities, the results from
\citet{Li2022} show that the spread in the [Mg/Ca] ratio
increases as [Fe/H] decreases with values ranging from negative
abundance ratios to highly enhanced [Mg/Ca] ratios, also suggested by 
\citet{francois_detailed_2020}.
 The result for SDSSJ081554.26$+$472947.5 found by
\citet{gonzalez_hernandez_extreme_2020} seems to corroborate this
hypothesis. The increase of the [Mg/Ca] ratio is also visible when the
ratio is plotted as a function of [Mg/H], as shown in the lower panel of Fig. \ref{fig:MgCa}. However, a high value
of [Mg/Ca] at [Mg/H] $\lesssim -4$ is not found in all the stars. In
particular, SDSS J102915+17292, the carbon-normal UMP star \citep{caffau_extremely_2011}, has a solar [Mg/Ca]
ratio. Our new results are not confirming this increase of the [Mg/Ca] ratio at low metallicity.
All these results support the hypothesis of an increasing dispersion in the [Mg/Ca] ratio at very low metallicity, likely resulting from the inhomogeneous chemical enrichment during the early evolution of the Galaxy.
When the [Mg/Ca] is plotted against [Mg/H], we find 
a large [Mg/Ca] dispersion of more than 1 dex for the  same  [Mg/H]  $\simeq -4.0$ dex.   
It is important to note that the dispersion is also influenced by the presence of additional Mg-poor stars, as these stars are not Ca-rich.
However, we should remind that the resonance line  we used for the determination of the Ca LTE abundance 
leads to an  underestimation of the calcium abundance at a level of $+0.1$ dex for turn-off stars as shown by 
 \citet{spite_nlte_2012}.

We also compared stars in the Milky Way with those in dwarf galaxies in the local group.
Stars were selected from the SAGA database with a metallicity cut at [Fe/H] $= -1.5$.
While the sample includes 601 stars from 26 dwarf galaxies, the abundance dispersion in these systems is notably larger than in the Milky Way and increases further toward higher metallicity, up to [Fe/H] $\sim 0$.
In contrast, the dispersion among Galactic stars remains relatively small across the range [Fe/H] $\gtrsim -3$, based on over 5000 stars compiled in the SAGA database.
This difference may reflect the effects of inhomogeneous chemical enrichment, which appears to persist in dwarf galaxies but has largely been smoothed out in the Milky Way.

\subsection{Neutron capture elements}

\begin{figure}
 \begin{center}
\includegraphics[width=8cm]{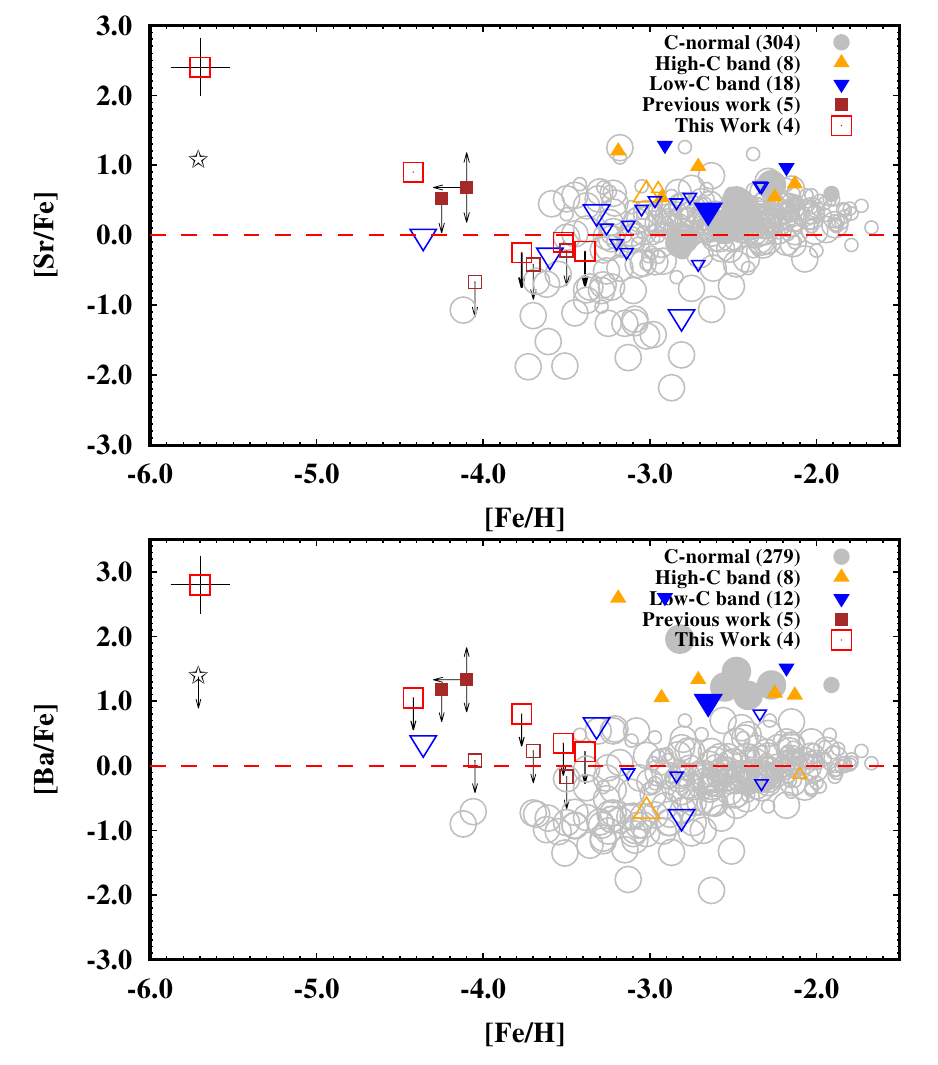} 
 \end{center}
\caption{[Sr/Fe] and [Ba/Fe] as a function of  [Fe/H]. The meaning of the symbols is the same as in Fig.~\ref{fig:alpha}. The star symbol represents HE1327-2326, which exhibits an exceptionally high [Sr/Fe] ratio \citep{frebel_he_2008, aoki_he_2006}.
{Alt text: Two dot graphs showing [Sr/Fe] and [Ba/Fe] as a function of [Fe/H].}
}
\label{fig:ncapture}
\end{figure}

\begin{figure}
 \begin{center}
\includegraphics[width=8cm]{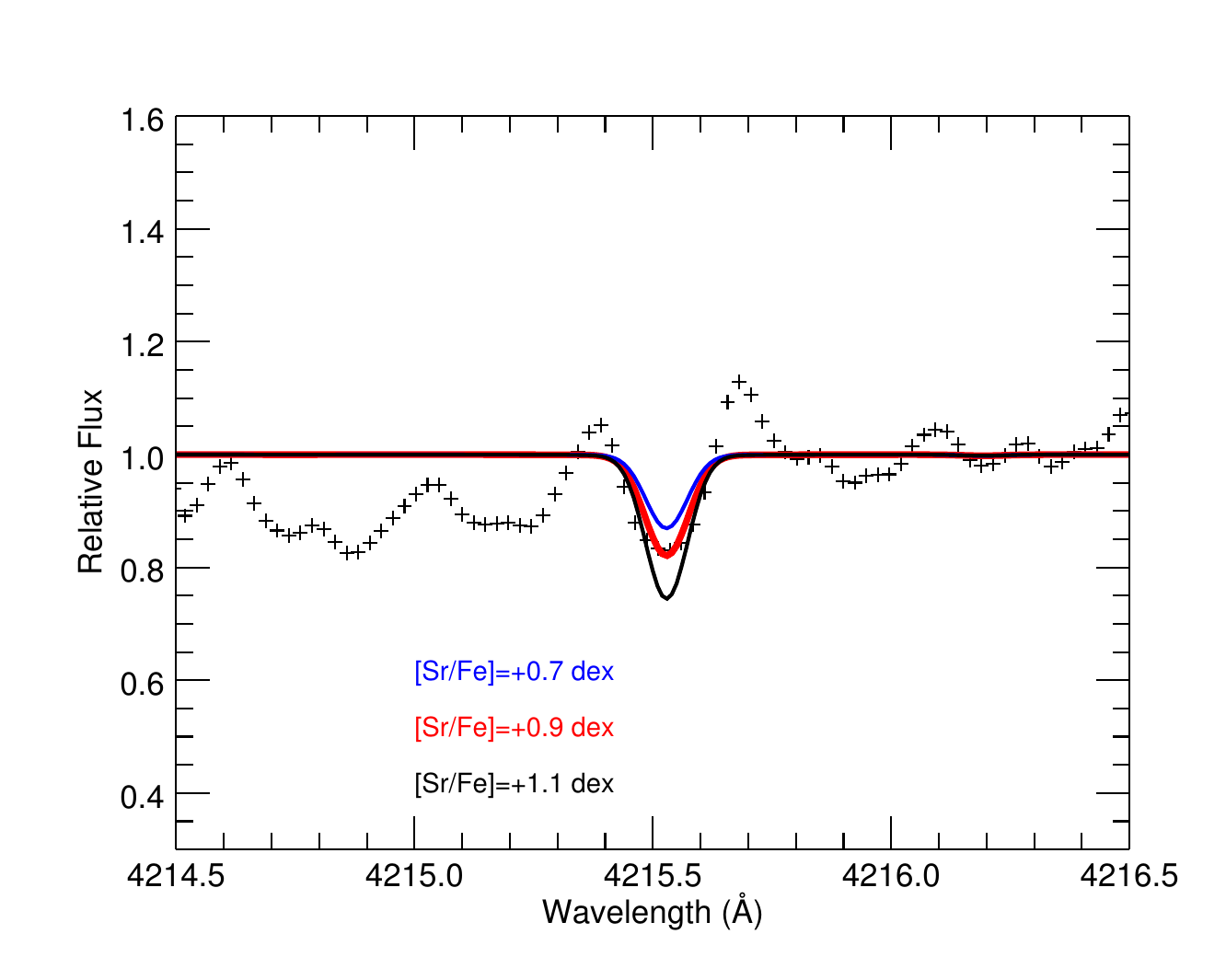} 
 \end{center}
\caption{ Comparison of the observed spectrum of SDSJ161956.33$+$170539.9   with  synthetic spectra in the region of the  421.5 nm Sr line. The observed spectrum is shown as '+' symbols. The theoretical spectra are represented as solid lines.
{Alt text: Three line graph with dots showing the fit of a spectral synthesis models to the observed spectrum.}
}
\label{fig:strontium_fit}
\end{figure}

Figure \ref{fig:ncapture} shows the  abundance ratios  of the two neutron capture elements  strontium and barium as a function metallicity.
Results from \citet{Li2022} are added as representative of a large homogeneous sample of evolved and unevolved stars.
For two stars, we measured the abundance of strontium.
Only upper limits could be determined for barium. Low abundances or undetermined values of neutron-capture elements are indicated by open symbols, as shown in Figure \ref{fig:carbon}.
While the strontium abundance in the star SDSS~J095932.52$+$265358.6 is slightly underabundant, we found a very high abundance of strontium in SDSS~J161956.33$+$170539.9, the most metal poor star of our sample with a ratio [Sr/Fe] $=+0.9$ dex.
Fig. \ref{fig:strontium_fit} show the comparison between the observations and  the theoretical spectra for  three different values of the strontium abundance.
A high strontium abundance has also been found in HE 1327-2326 \citep{frebel_he_2008, aoki_he_2006}.
Further observations are needed to confirm this result and would be useful in obtaining a detection or at least a significant upper limit for Ba.
In one of the two low n-capture stars, we have a detection of Sr with [Sr/Fe] $= -0.10$ dex  similar to the values found in giant stars from \citet{Li2022}. This almost solar value of [Sr/Fe] is not found in the main sequence sample of \citet{Li2022}. Our results seem to show that metal-poor giants and main sequence stars both exhibit an increasing dispersion as metallicity decreases. Increasing the sample of  Sr and Ba measurements in main sequence stars in the metallicity range of $-2.5$ to $-3.0$ would be useful to estimate this dispersion and compare it with  what is  found for the giant sample. At lower metallicity, the measurement of their abundance remains challenging. 

\section{Summary}
In this article, we reported the chemical analysis of four
extremely metal-poor candidates observed with the high dispersion spectrograph (HDS) at the Subaru telescope.
None of these stars has been studied at high resolution so far. Their selection was made on the analysis of SDSS spectra 
using the  method created by \citet{ludwig_extremely_2008} to estimate the metallicity of
turn-off stars from low-resolution spectra. 
Among these stars, we discovered a new UMP star SDSSJ61956.33$+$170539.9 with a metallicity [Fe/H] $= -4.42$, which is the second-most iron-poor star discovered by the Subaru telescope.
Given that only two out of 14 stars with [Fe/H] $< -4.5$ are discovered so far, more such stars can be found using large ground-based telescopes in the Northern Hemisphere.
We were able to determine the abundances of some elements (C, Mg, Ca, Sr and Ba) in the majority of these stars.
We measure a lithium abundance A(Li) = 2.1 dex in the star SDSS~J125601.88$+$460836.8 of metallicity [Fe/H] $= -3.39$, where the lithium 
melt-down starts to operate. Two stars of the sample show abundance ratios which are typical of metal-poor stars 
in the metallicity range $-4.42$ dex $\le$ [Fe/H] $\le -3.39$ dex, whereas the two other show low [$\alpha$/Fe] ratios confirming the presence
of low [$\alpha$/Fe] stars at low metallicity. 
Remarkably one of these two stars, SDSS~J125601.88$+$460836.8, is likely associated to the Sequoia/Thamnos accretion event.
We  measured a very high [Sr/Fe] ratio = +0.9 dex in  the new UMP star SDSSJ61956.33$+$170539.9, a similar value also found in the star HE 1327-2326, which belongs to the group of the most iron-poor stars known.


\begin{ack}
We thank the anonymous referee for carefully reading the manuscript and providing valuable comments.
T. S. acknowledges support by a Grant-in-Aid for Scientific Research (KAKENHI) (JP21H04499,JP22K03688, JP23HP8014, JP24HP8010) from the Japan Society for the Promotion of Science (JSPS).
T. S. and W. A also acknowlege support from JSPS KAKENHI (JP25K01046).
P.F. acknowledges support from the AIPS and the GGS of the Paris Observatory.
PB acknowledges support from the ERC advanced grant No.835087 -- SPIAKID.
\end{ack}

\bibliographystyle{apj} 
\bibliography{EMP_Subaru_2024}

\appendix 
\section*{Dependence on calcium resonance line}
\begin{figure}
\includegraphics[width=8cm]{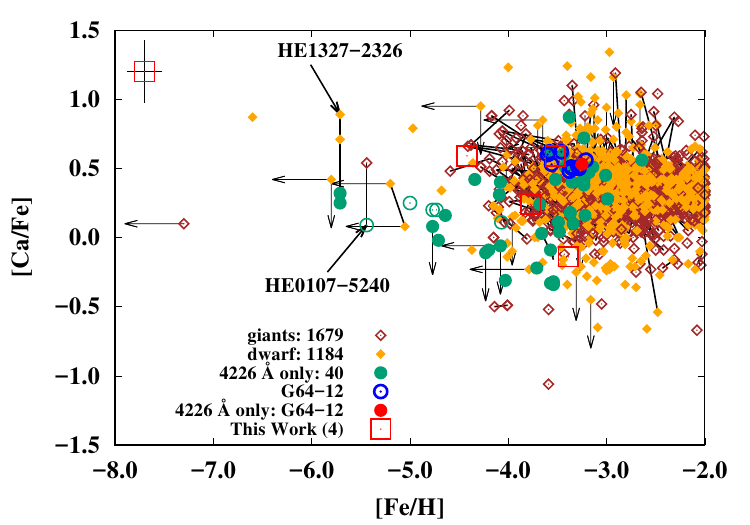} 
\caption{Comparison of calcium abundances derived from the resonance line at 4226 \AA\ and from other lines, such as the Ca II H and K lines, the triplet, and combinations of multiple lines.
Green circles represent stars with Ca I abundances derived from the resonance line; filled and open circles correspond to dwarfs and giants, respectively.
Open and filled diamonds indicate giants and dwarfs, respectively, whose calcium abundances were determined without using the resonance line, were based on a mix of lines, or for which the line information is not available.
Abundances for the same stars are connected by solid lines.
The filled red circle represents G~64-12 with the resonance line, while the open blue circle corresponds to G~64-12 without it.
Red open squares mark the stars analyzed in this study, for which Ca abundances were derived from the resonance line.
For HE~1327-2326 and HE~0107-5240, calcium abundances based on both the resonance line and other lines are plotted for comparison.
{Alt text: Scatter plot showing calcium abundances as a function of metallicity, comparing results from different spectral lines.}
}
\label{fig:cafe}
\end{figure}

Calcium abundances are known to depend on the choice of absorption lines used in the analysis. In particular, abundances derived from the Ca resonance line at 4226 \AA\ tend to be lower than those obtained from the Ca II H and K lines at 3933 and 3969 \AA, as well as from the infrared Ca II triplet lines.
Figure~\ref{fig:cafe} compares calcium abundances derived from the resonance line with those obtained using other methods.
The discrepancy becomes more pronounced at lower metallicities, as seen in the cases of HE~1327-2326 and HE~0107-5240, whereas no significant difference is observed for G~64-12.

\bigskip
\noindent

\end{document}